\documentclass[sigconf]{acmart}

\usepackage{graphicx}
   \graphicspath{{./Figures/}}

\usepackage{amsmath}    
\usepackage{algorithm}
\usepackage{algorithmic}
\usepackage{multirow}
\usepackage{svg}
\usepackage{color}
\DeclareGraphicsExtensions{.pdf,.jpeg,.png,.fig, .emf}
   \usepackage{subfigure}
    \usepackage{epstopdf}
    \usepackage{epsfig}
    \usepackage{subfigure}
    \usepackage{epstopdf}
    \usepackage{epsfig}

\newcommand{\cosi}{COSIME}
\copyrightyear{2022}
\acmYear{2022}
\setcopyright{acmcopyright}\acmConference[ICCAD '22]{IEEE/ACM International
Conference on Computer-Aided Design}{October 30-November 3, 2022}{San Diego, CA, USA}
\acmBooktitle{IEEE/ACM International Conference on Computer-Aided Design (ICCAD
'22), October 30-November 3, 2022, San Diego, CA, USA}
\acmPrice{15.00}
\acmDOI{10.1145/3508352.3549412}
\acmISBN{978-1-4503-9217-4/22/10}
\makeatletter
\let\ACM@origbaselinestretch\baselinestretch
\makeatother
\begin{document}
%
\title{COSIME: FeFET based Associative Memory for In-Memory Cosine Similarity Search}

 \author{\small{}
         Che-Kai~Liu$^{1,2}$, Haobang~Chen$^1$, Mohsen~Imani$^3$, Kai~Ni$^4$, Arman~Kazemi$^2$, Ann~Franchesca~Laguna$^5$, Michael~Niemier$^2$, Xiaobo~Sharon~Hu$^2$, Liang~Zhao$^1$, Cheng~Zhuo$^{1,*}$, and Xunzhao~Yin$^{1,6,*}$\\
    $^1$College of Information Science and Electronic Engineering, Zhejiang University, Hangzhou, China\\
    $^2$Department of Computer Science and Engineering, University of Notre Dame, IN, USA\\
    $^3$Department of Computer Science, University of California, Irvine, CA, USA\\
    $^4$Department of Electrical and Microelectronic Engineering, Rochester Institute of Technology, NY, USA\\
    $^5$Department of Computer Technology, De La Salle University, Manilla, Philippines\\
    $^6$Zhejiang Lab, China\ $^*$Corresponding authors, email: \{czhuo, xzyin1\}@zju.edu.cn
}

\begin{abstract}


In a number of machine learning models, an input query is searched across the trained class vectors to find the closest feature class vector in cosine similarity metric. However, performing the cosine similarities between the vectors in  Von-Neumann machines involves a large number of multiplications, Euclidean normalizations and division operations, thus incurring heavy hardware energy and latency overheads. 
Moreover, due to the memory wall problem that presents in the conventional architecture, frequent cosine similarity-based searches (CSSs) over the class vectors requires a lot of data movements, limiting the throughput and efficiency of the system.
To overcome the aforementioned challenges, this paper introduces {\cosi}, an general in-memory associative memory (AM) engine based on the ferroelectric FET (FeFET) device for efficient CSS.
By leveraging the one-transistor AND gate function of FeFET devices, current-based translinear analog circuit and winner-take-all (WTA) circuitry, {\cosi} can realize parallel in-memory CSS across all the entries in a  memory block, and output the closest word to the input query in cosine similarity metric.
Evaluation results at the array level suggest that the proposed {\cosi} design achieves 333$\times$ and 90.5$\times$ latency and energy improvements, respectively, and realizes better classification accuracy when compared with an AM design implementing approximated CSS.  
The proposed in-memory computing fabric is evaluated for an HDC problem, showcasing that {\cosi} can achieve on average 47.1$\times$ and 98.5$\times$ speedup and energy efficiency improvements compared with an GPU implementation.

\end{abstract}


\maketitle
\pagestyle{empty}

%



\section{Introduction}
\label{sec:introduction}

Cosine similarity measures the similarity between two vectors in an inner product space. It is widely used in a number of machine learning models such as hyperdimensional computing (HDC) and deep neural networks (DNNs).
During the inference phase of these machine learning applications, a large number of cosine similarity-based searches (CSSs) are often needed.
While CSS has been extensively studied in many algorithm-level approaches~\cite{Gong_2018}, and can be executed in digital machines, it requires a large number of multiplications, as well as $L_2$ normalizations and divisions. 
Moreover, given the extensive search operations required by many machine learning algorithms such as HDC classifications, CSS 
also causes massive data movements between the memory and the processing units, i.e. the memory wall problem, limiting its performance and efficiency.
These challenges posed by CSS becomes even more significant when deployed in power and resource constrained scenarios~\cite{area_2018}, calling for innovative and efficient hardware design for CSS.

Compute-in-memory (CiM) is a promising architectural paradigm that enables  operations across entire memory blocks by integrating some basic processing capabilities inside the memory to overcome the memory wall problem. 
For example, content addressable memories (CAMs)~\cite{hu2021memory,amrouch2021iccad}, which support parallel search across the stored vectors in memory against the input query, have been proposed as associative memories (AMs) to accelerate inference for machine learning applications e.g., few-shot learning, transformer, etc.,~\cite{CAM_2015, Ni_2019}. 
Moreover, in conjunction with non-volatile memory (NVM) technologies such as resistive RAM (ReRAM), ferroelectric FET (FeFET), etc., and customized sense amplifier (SA), CAM designs have demonstrated great potential as highly energy efficient AMs for nearest neighbor (NN) search using the Hamming distance metric \cite{Ni_2019, accuracy, kazemi2021memory, Imani_2017}.

\begin{figure}
    \centering
    \includegraphics[width=\linewidth]{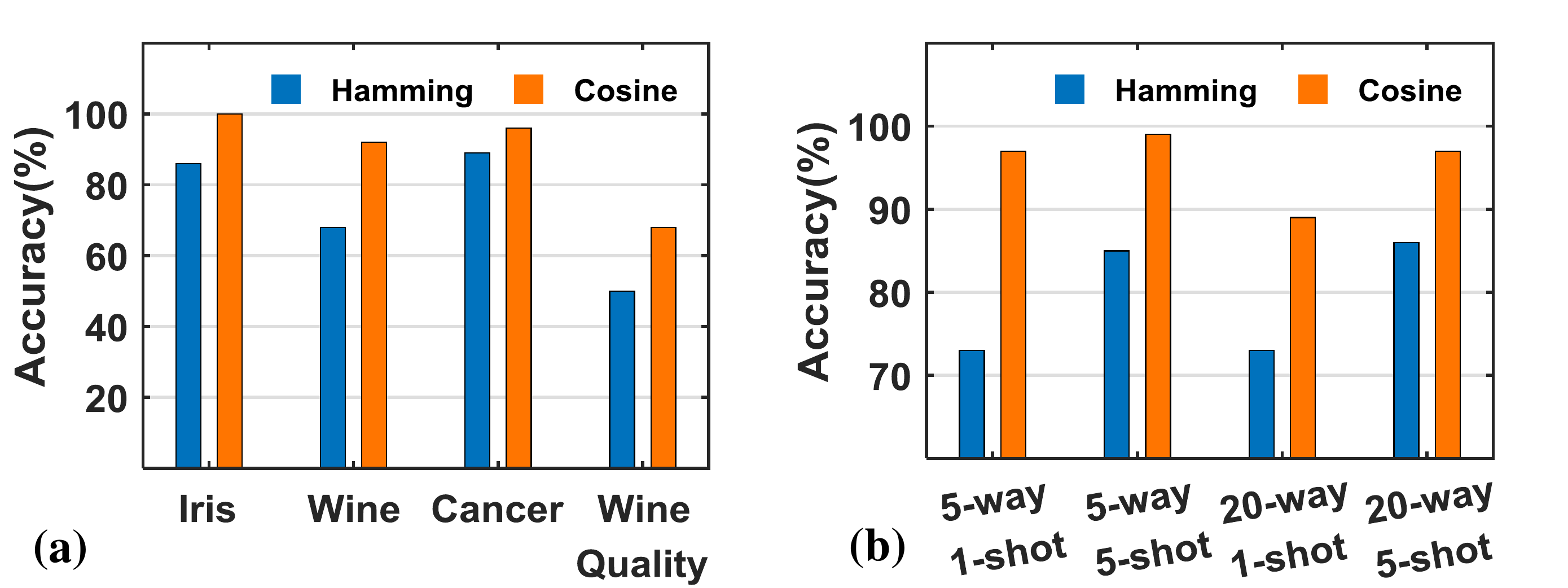}
    \caption{ Accuracy of (a) nearest neighbor classification, (b) few-shot learning tasks with Hamming distance based search and CSS \cite{accuracy}, respectively.}
\label{fig:accuracy}
\end{figure}

However, it can be seen in Fig. \ref{fig:accuracy} that the CAM designs supporting Hamming distance based search achieves energy efficiency at the expense of non-negligible accuracy loss for classification tasks.
The AM in \cite{Geethan2021} supports a specific approximated CSS by approximating the denominator of cosine calculation and exploiting the the quasi-orthogonal property of hyperdimensional vector. 
That said, such approximation still causes slight accuracy loss, and is limited only to the HDC application. 
Therefore, an more general CiM based CSS that can not only offer energy efficiency and performance improvements but also maintain comparable accuracy to the full precision CSS implemented in software is highly desirable.

In this paper, we propose {\cosi}, an energy efficient, FeFET based in-memory search engine that implements parallel CSS across the memory to identify the word closest to the input query. 
{\cosi} incorporates several novel circuit designs summarized below.
FeFETs are exploited as non-volatile storage to store the  pre-trained data words (e.g., pre-trained class vectors in machine learning applications) in two FeFET arrays, where one array enables a row-wise dot product calculation between the input query and all stored words and the other array implements bit counting of the stored vectors. 
Current-mode analog circuits are employed to efficiently realize the cosine similarity calculation as well as the NN search. 
Since the analog circuits are independent of the memory array,  the proposed {\cosi} design is not limited to FeFET technology, but can also be applied to other NVMs with access transistors.

To validate the functionality and evaluate the scalability and robustness of {\cosi}, NN searches based on cosine similarity are performed and analyzed.
Evaluation results suggest that {\cosi} achieves 
$90.5\times$ energy saving and $333\times$ latency reduction compared to 
the AM design that implements approximated CSS~\cite{Geethan2021}. 
To benchmark {\cosi} at the application level, we use {\cosi} for HDC classification where {\cosi} is implemented as an inference-accelerating AM. In this setup, {\cosi} demonstrates 47.1$\times$ and 98.5$\times$ speedup and energy efficiency improvement while maintaining the same classification accuracy compared with a GPU. 
To the best of our knowledge, {\cosi} is the first CiM design that supports NN search based on accurate cosine similarity.

\section{Background}
\label{sec:background}
In this section, we review FeFET basics and justify why FeFET is a favorable choice for implementing CiM based CSS. We then summarize recent efforts on designing AMs for similarity search.

\subsection{FeFET Basics}
\label{sec:device}
\begin{figure}
    \centering
    \includegraphics[width=1\linewidth]{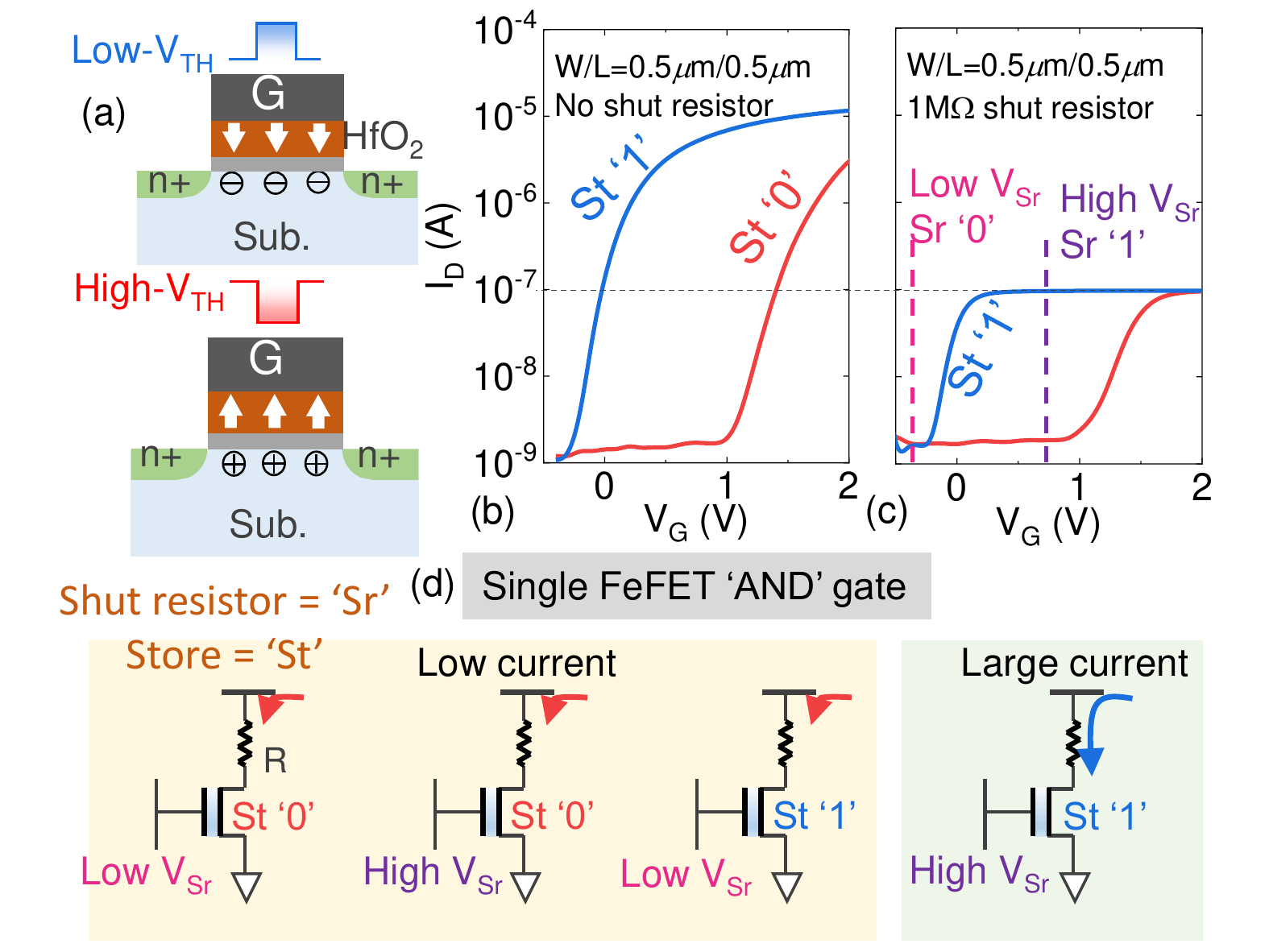}
    \caption{(a) FeFET operation principles. $I_D$-$V_G$ characteristics of the high-$V_{TH}$ and low-$V_{TH}$ states for (b) a single FeFET and (c) a FeFET with a series resistor on the drain. (d) A single FeFET can compactly realize the AND gate.}
    \label{fig:fefet}
\end{figure}

FeFET based on recently discovered ferroelectric HfO$_2$ is a competitive candidate for high-speed, high-density, and low-power embedded NVM due to its intrinsic transistor structure, CMOS compatibility, excellent scalability, and
superior energy efficiency \cite{boscke2011ferroelectricity}. 
As shown in Fig.~\ref{fig:fefet}(a), 
applying a positive (negative) gate voltage pulse
sets the FeFET to low-$V_{TH}$ (high-$V_{TH}$) state (Fig.~\ref{fig:fefet}(b)).
Unlike others NVMs whose memory write is driven by current and consumes significant power, FeFET exhibits superior write energy efficiency since the polarization switching is driven by an electric field, rather than large conduction currents \cite{Ni_2019}.

It is demonstrated in \cite{soliman2020ultra} that by connecting a series resistor with proper resistance value on the FeFET source/drain, the ON state current will be only limited by the series resistance, as shown in Fig.~\ref{fig:fefet}(c).  
As a result, the ON state current variation is significantly reduced with such 1FeFET1R structure and made independent from the FeFET $V_{TH}$ variation. This suggests possible tuning to the FeFET ON current for both the low-$V_{TH}$ and high-$V_{TH}$ states.
\cite{1F1R_2021} experimentally demonstrated a back-end-of-line (BEOL) 1FeFET1R structure, 
validating the aforementioned ON current tuning scheme with smaller cell area than other devices. 
A resistor with less than 8\% variability is demonstrated. 
Given the small 1R variability and relatively large resistance of R, the ON state current of 1FeFET1R cell is approximately proportional to $\frac{1}{R}$ due to the large $M\Omega$ resistance of 1R, and the ON state current variation $\Delta I$, i.e., the derivation of ON state current, is proportional to $-\frac{\Delta R}{R\times R}$, where $\frac{\Delta R}{R}$ refers to the 1R variability.
Therefore,
the impact of the 1R variability on the ON state current is negligible \cite{soliman2020ultra}.

In this work, the 1FeFET1R structure is adopted, 
and proposed to realize a compact AND gate (i.e., dot product for binary vectors) by storing one operand as the FeFET $V_{TH}$ state and applying the other operand as the gate voltage, as shown in Fig.~\ref{fig:fefet}(d) \cite{Xunzhao_2019}. Such cell is leveraged in this work to calculate the cosine similarity in-memory.

    

\vspace{-1ex}
\subsection{Existing Associative Memory with Similarity Search}
\label{sec:existing_work}

With advances in NVMs, binary/ternary/multi-bit CAM design have been proposed for energy efficient and ultra-dense associative search in various applications, e.g., IP routers, look-up table, reconfigurable computing and machine learning models, etc. \cite{kohonen2012associative, yin2020fecam, li2020a, rajaei2020compact, laguna2020seed}.
Typically, CAM works in the exact match mode, in which only the stored vectors that exactly match the query are identified.
However, NN search is also highly desirable as it identifies the vector closest to the query, a core computation in many machine learning models.
With the exact matching mode, to identify the NN, multi-step searches with queries of increasing distance to the target query are applied, incurring overheads in energy and latency. 

This can be overcome by leveraging the approximate matching mode of CAM, which directly computes the Hamming distance on the match line (ML) of CAM. Recent work \cite{Ni_2019, accuracy, kazemi2021memory, Imani_2017} implemented NN search based on Hamming distance for few-shot learning tasks. 
However, they suffer from significant accuracy loss compared with CSS, as shown in Fig. \ref{fig:accuracy}.
Recently, an AM supporting approximate CSS 
was proposed in \cite{Geethan2021}. This design specifically targets to HDC classification problems and approximates the denominator of cosine calculation by exploiting the quasi-orthogonal property of hyperdimensional vectors, thus limiting its application to other machine learning models.
In this work, we propose a more general AM design that supports NN search based on cosine similarity.

\begin{figure*}
    \centering
    \includegraphics[width=1\linewidth]{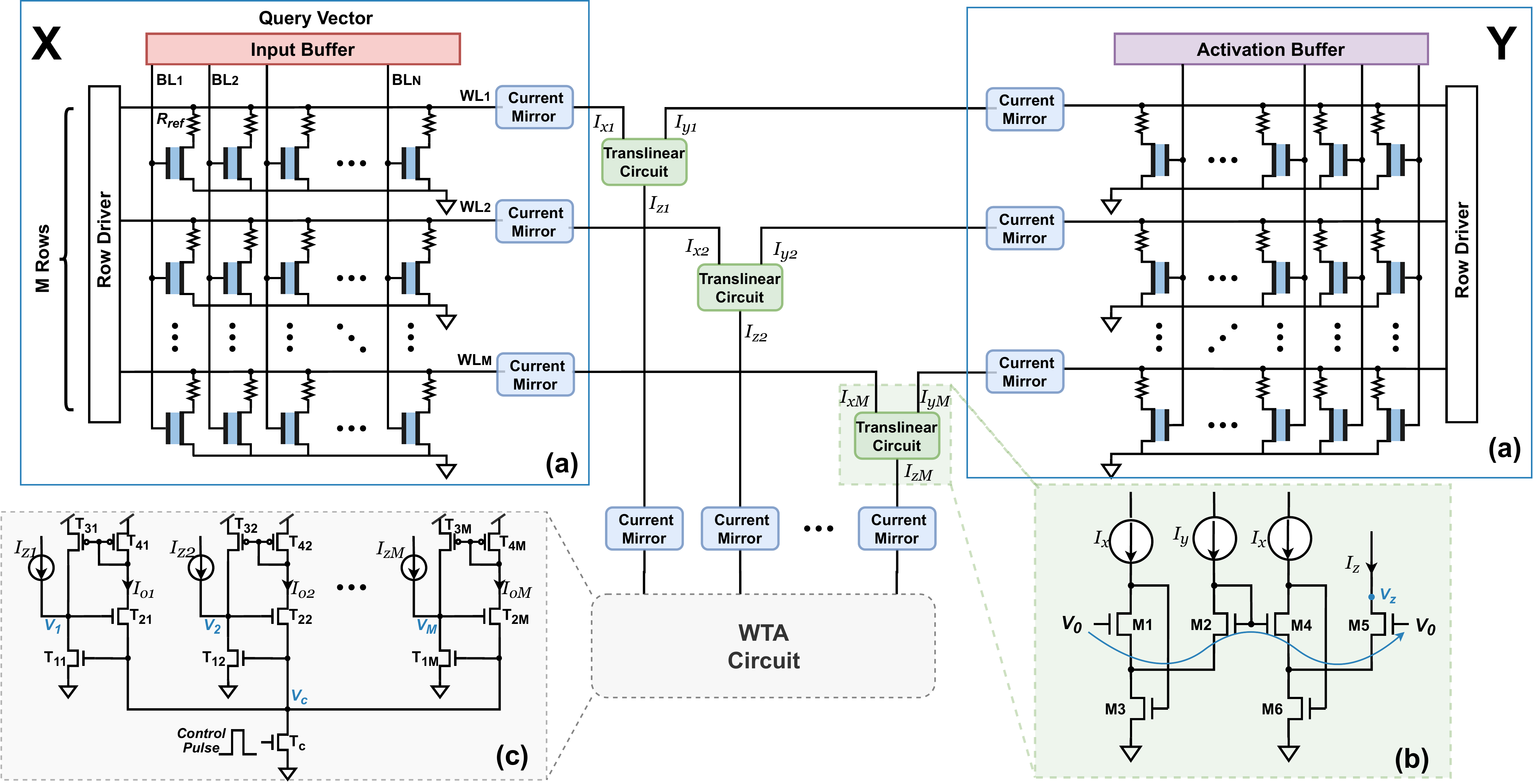}
    \caption{{\cosi} overview. (a) 1FeFET1R memory array. (b) Translinear circuit. (c) Winner-Take-All (WTA) circuit.}
    \label{fig:overall}
\end{figure*}

\section{{\cosi}: In-memory Cosine similarity search engine}
\label{sec:architecture}

{\cosi} implements in-memory cosine similarity search for the NN, a critical operation in the inference phase of BNN and HDC models as well as other machine learning models (e.g., for few-shot learning). 
Fig.~\ref{fig:overall} shows the architecture of {\cosi}.  
It consists of two FeFET memory arrays, a current-mode translinear circuit block for each row in the memory arrays, and an analog winner-take-all (WTA) circuit.
Below we elaborate the cosine similarity metric derivation, as well as the detailed design of the {\cosi} components.

\subsection{Cosine Similarity}
\label{sec:cosine_derivation}

Cosine similarity (or cosine distance) has been used as a distance metric for measuring the difference between an input feature query vector and the stored vectors. 
Specifically,
\begin{equation}
\label{eq:cosine}
    \cos \langle \vec{a}, \vec{b} \rangle = \dfrac{\vec{a}\cdot \vec{b}}{\lVert \vec{a} \rVert \times \lVert \vec{b} \rVert}
\end{equation}
Without loss of generality, in this work we assume that $\vec{a}$ is the binary input vector whose bits are either 0 or 1, and $\vec{b}$ as the  class binary vector stored in the memory block. 
Eq. \eqref{eq:cosine} equals to 1 means that $\vec{a}$ and $\vec{b}$ are exactly the same, while Eq. \eqref{eq:cosine} equals to 0 means that $\vec{a}$ and $\vec{b}$ are orthogonal to each other.
The numerator of Eq. \eqref{eq:cosine} is simply the dot product of two vectors.  
The denominator is the product of the $L_2$ norm of the two vectors, while $L_2$ norm of a binary vector is the square root of the number of `1's in the vector. 
Directly computing $L_2$ norm requires a complex circuit, hindering the denominator from efficient hardware implementation. 
Therefore, prior works typically simplify the cosine equation by removing the denominator, or approximating the denominator to a constant value. Doing so may introduce significant errors.

{\cosi} aims to obtain the closest vector  (i.e., NN) to the input query in terms of cosine similarity. 
From Eq.~\eqref{eq:cosine}, cosine similarity can be equivalently expressed in a more circuit friendly variant without affecting the search output. 
Specifically, for computing cosine similarity, {\cosi} removes the need for square root operation without any accuracy loss by squaring both the numerator and denominator as shown in Eq. \eqref{eq:square}.
\begin{equation}
\label{eq:square}
\cos^2\langle \vec{a}, \vec{b} \rangle = \dfrac{(\vec{a}\cdot\vec{b})^2}{( \lVert\vec{a}\rVert \times \lVert \vec{b} \rVert)^2}
\end{equation}
Note that in Eq. \eqref{eq:square}, the denominator consists of the squared norm of the stored vector $\vec{b}$ which is the number of `1's within $\vec{b}$, and squared norm of input query vector $\vec{a}$ which is shared by all the cosine similarity metrics, and thus can be removed during the CSS. 
In this sense, the cosine similarity metric can be equivalently expressed as the $X^2/Y$ operator, where $X$ denotes the dot product $(\vec{a}\cdot\vec{b})^2$, and $Y$ denotes $\lVert \vec{b} \rVert^2$, i.e., the number of `1's within $\vec{b}$.
Based on the above formulation, we illustrate the circuits implementing the computation of $X$, $Y$ and $X^2/Y$.

\subsection{FeFET Memory arrays}
\label{sec:mem_array}
We employ two identical FeFET-based non-volatile memory arrays to store the class vectors.
As shown in Fig. \ref{fig:overall}(a),  the gates of the FeFETs within a column are connected to the bitlines (BLs), while the drains of the FeFETs within a word share a wordline (WL). 
As discussed in Sec. \ref{sec:device}, FeFETs can be used as a single transistor AND gate, enabling the memory array implementing in-memory binary bitwise dot product naturally.

During search, for the FeFET memory array on the left side of Fig. \ref{fig:overall}(a), high/low voltages are applied to the bitlines $BL$ according to the respective bit values in the input query. Only when the FeFET of a cell stores `1' (corresponding to low $V_{TH}$ state), and its gate voltage is at high level indicating an input bit `1', the cell is turned on, conducting $I_{ON}$ current from the wordline WL to ground.
The resulting output current ($I_x$) flowing through a WL therefore represents the dot product of this word and the input query, i.e., $X$.
To implement  the squared norm of the stored vector, i.e., $Y$, the FeFET memory array on the right side of Fig. ~\ref{fig:overall}(a) is used, and stores the identical class vectors as the left memory array.
All the bitlines of this array, however, are applied with high gate voltage, turning on the FeFETs storing `1'. It is easy to see that the magnitude of the output current $I_y$ of a word in this array represents the number of `1's within the stored vector, i.e., $Y$.
Note that the magnitude of the output currents $I_x$, $I_y$ can be adjusted by tuning the resistor within the 1FeFET1R structure as discussed in Sec. \ref{sec:device}, ensuring that the input currents are in the working range of the following translinear circuit model.

\subsection{Translinear Circuits}
\label{sec:translinear}
To implement the key operation $X^2/Y$ for CSS, we propose to employ the translinear circuit from \cite{Minch_2009} and feed the output currents of FeFET memory arrays $I_x$ and $I_y$ into this analog arithmetic circuit. 
Fig. \ref{fig:overall}(b) shows the schematic of the translinear circuit implementing efficient current-mode squaring and division.
This translinear circuit mainly consists of a translinear loop (indicated by the blue arrow) including clockwise (CW) transistors M1, M4 and counter-clockwise (CCW) transistors M2, M5. 
The transistors along the loop are operating in the subthreshold (weak inversion) region, and their drain-source currents can be characterized by the following expression \cite{ANDREOU_1996}:
\begin{equation}
\label{eq:ids}
I_{DS} \approx I_{0}\frac{W}{L}e^\frac{V_{GS}}{\eta V_T}
\end{equation}
where $I_0$ denotes the drain current $I_D$ when $V_{GS} = V_T$, $V_T$ denotes the thermal voltage, $\eta$ the subthreshold slope factor.

The relation between the $V_{GS}$'s of the transistors  along the translinear loop follows Kirchoff's Law, i.e.,
\begin{equation}
\label{eq:kirchoff}
    \sum_{CW}V_{GS} = \sum_{CCW}V_{GS}
\end{equation}
from Eq. \eqref{eq:ids}, we obtain:
\begin{equation}
\label{eq:vgs}
    V_{GS} = V_T\eta ln(\frac{I_{DS}}{I_0})
\end{equation}
By substituting Eq. \eqref{eq:vgs} into Eq. \eqref{eq:kirchoff} while keeping the loop transistors in the subthreshold region, 
the translinear circuit generates the analog output current $I_z$ as below:
\begin{equation}
\label{eq:translinear}
    I_z=\frac{I_x^2}{I_y}
\end{equation}

\begin{figure}
    \centering
     \includegraphics[width=\linewidth]{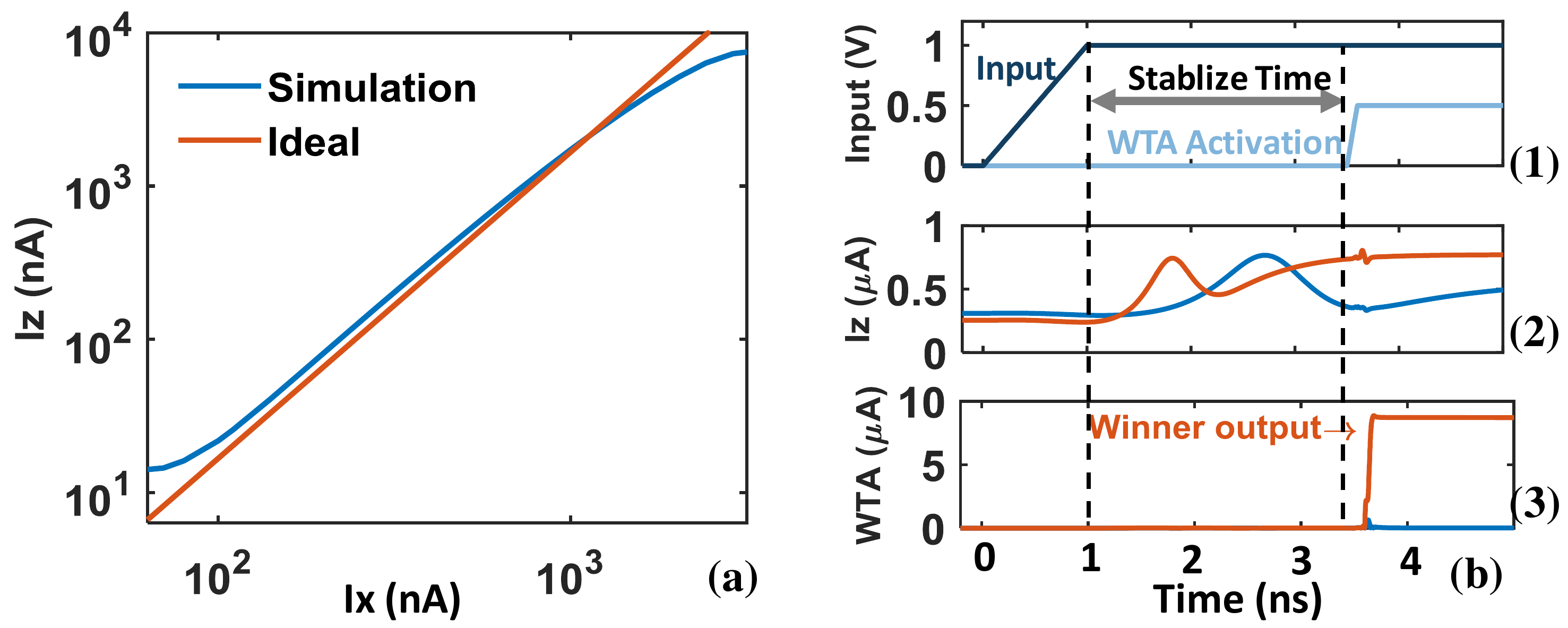}

    \caption{(a) The transfer characteristic of the translinear circuit, where the center linear region indicates the operating region for the input. (b) 
    Waveforms of (1) input and WTA activation for NN search, (2) translinear circuit output, and (3) WTA output.}
    \label{fig:comprehensive_translinear}
\end{figure}

The operating voltage $V_0$ in Fig. \ref{fig:overall}(b) is set to 0.6V to keep the translinear circuit in the subthreshold region, and $I_y$ around 600nA corresponding to the average squared $L_2$ norm of the stored vectors. 
Fig. \ref{fig:comprehensive_translinear}(a) shows the operating region with respect to the input current $I_x$, where the simulated transfer characteristic aligns with the theoretical result. 
It can be seen that the input current $I_x$ from the FeFET memory array should be within the operating range to guarantee the functionality of the translinear circuit. 
In order to maintain the correct functioning of the translinear circuit and guarantee the scalability of {\cosi}, we propose to adjust the resistor in every 1FeFET1R to satisfy the  required input current range of the translinear circuit. 
For example, when the memory arrays scale to $N$ times row-wise, the input current per row from the memory arrays can remain constant by tuning the 1FeFET1R structure as presented in \cite{soliman2020ultra}, thus reducing the 1FeFET1R cell ON current to $\frac{1}{N}$ times.
\begin{equation}
\label{eq:scale}
    I_z = \dfrac{(\frac{I_x}{N}\times N)^2}{\frac{I_y}{N}\times N} = \dfrac{I_x^2}{I_y}
\end{equation}
Moreover, the input current range can also be guaranteed by adjusting the size ratio $W/L$ of the current mirror associated with the translinear circuits. 



\subsection{Winner-Take-All Circuit}
\label{sec:wta}

By employing the FeFET memory arrays and translinear circuits discussed above, the squared cosine distances between the stored vectors and the input query are effectively represented by the output currents of the translinear circuits. 
The last stage of the NN search is to find the maximum current as the ultimate search result. 
The conventional maximum current selection implementation is a current comparator-based tree structure which requires a huge number of transistors and increases the latency as the number of stored class vectors increases \cite{Imani_2017}.
Here, we propose to utilize a current-mode O(N) WTA circuit presented in \cite{NEW_WTA}, which can offer efficient maximum current detection operation. 

Fig. \ref{fig:overall}(c) shows the schematic of the WTA circuit. 
It consists of a gated transistor $T_C$ as the current source, a coupled transistor pair (i.e., the sourcing transistor $T_{1i}$ and the output transistor $T_{2i}$) and an output feedback current mirror (i.e., $T_{3i}$ and $T_{4i}$) for each input and output branch.
The WTA circuit generally operates by inhibiting the transistor pairs sourcing smaller input currents, and amplifying the transistor pair sourcing the maximum input current.
When one of the input currents is larger than others, the gate voltage $V_c$ of the corresponding sourcing transistor is driven to a higher level, while the drain voltages of other sourcing transistors are driven to a lower level to maintain the smaller input currents.
As a result, the reduced drain voltages will drive less output currents through the output transistors, and the output transistor corresponding to the maximum input drives a larger output current. 
The feedback current mirror on the output path adds the output current back to the input path, thus further exacerbating the current differences among the inputs. Such a WTA circuit therefore can distinguish input currents with even 1\% difference.
To this end, the WTA realizes NN search in cosine similarity metric. 
Fig. \ref{fig:comprehensive_translinear}(b) shows the transient input and output waveforms of the WTA circuit.
Note that to ensure correct WTA functionality, the WTA is activated after the input currents, i.e. the translinar circuit outputs, become stable.
\subsection{Scalability of WTA Circuit}
\label{sec:wta_scale}
The  2-rail-input WTA circuit demonstrated in \cite{Imani_2017} may cost a large number of two-rail-input WTA components to construct a comparison tree. Instead, we hereby deploy an M-rail-input WTA circuit in \cite{WTA} to our {\cosi} design. In \cite{WTA}, the derivation 2-rail-input WTA's output w.r.t. the input changes is given. To understand M-rail-input WTA's counterpart,
below we elaborate why the transfer characteristics between the inputs and the outputs of WTA are weakly correlated with the number of input rail in {\cosi}, thus validating the scalability of {\cosi} by using the M-rail-input WTA circuit.

\begin{figure}
    \centering
    \includegraphics[width=\linewidth]{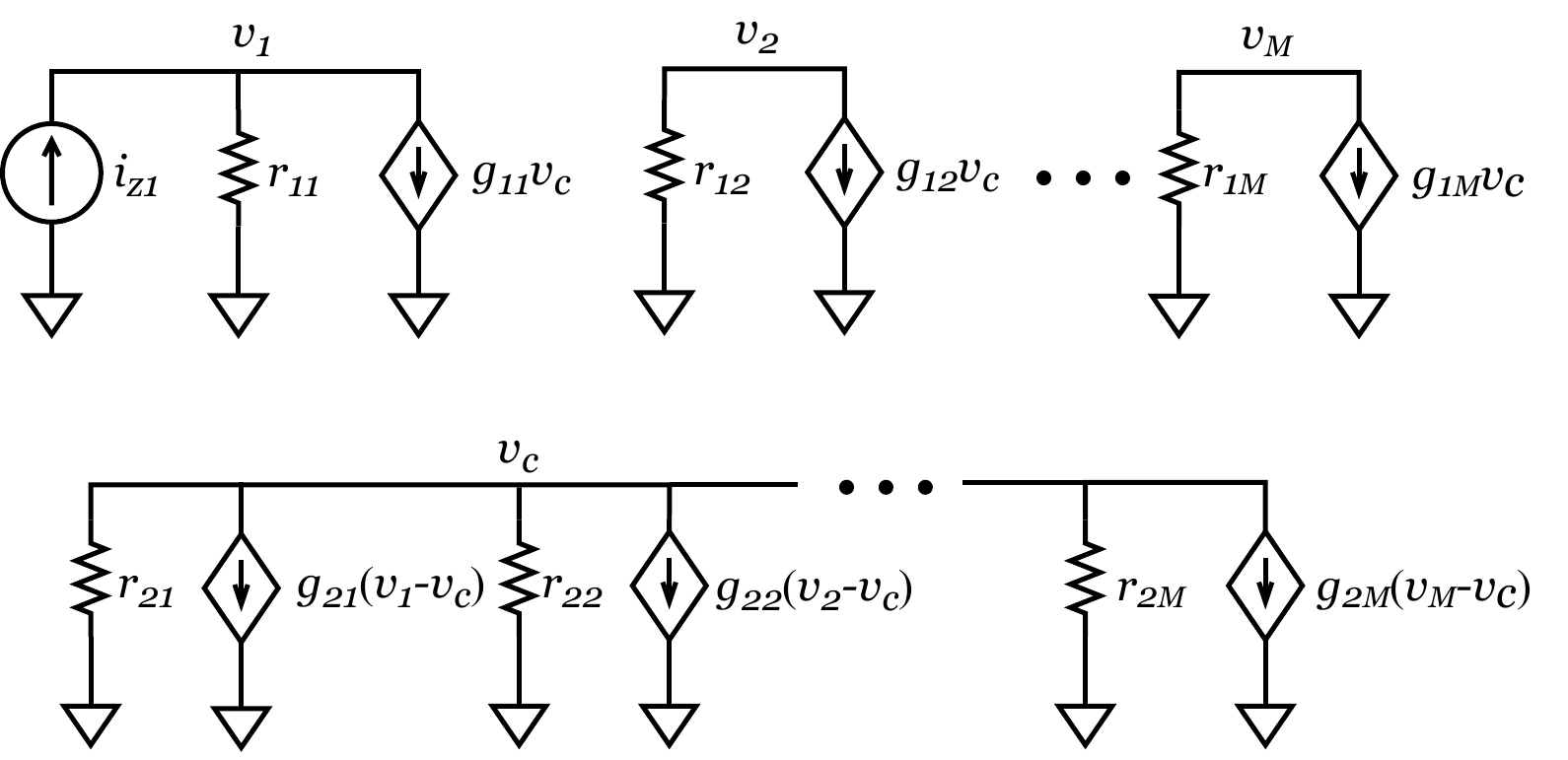}
    \caption{Small-signal model of the M-rail-input WTA circuit.}
    \label{fig:small_signal}
\end{figure}

Fig.~\ref{fig:small_signal} shows the small-signal circuit model of the M-rail-input WTA circuit excluding the feedback current mirrors. 
The small signal of $V_1$, $V_2$, and $V_c$ are denoted as $v_1$, $v_2$, and $v_c$, respectively.
For a particular operating point $[I_{z1},\dots,I_{zM},I_{o1},\dots,I_{oM}]$, 
without loss of generality, we assume a small change in $I_{z1}$, denoted as $i_{z1}$.  
the corresponding small-signal parameters of the sourcing and output transistors in the subthreshold region are $g_{11}=\frac{I_{z1}}{V_T}\dots g_{1M}=\frac{I_{zM}}{V_T}$, $g_{21}=\frac{I_{o1}}{V_T}\dots g_{2M}=\frac{I_{oM}}{V_T}$, $r_{11}=\frac{V_A}{I_{z1}}\dots r_{1M}=\frac{V_A}{I_{zM}}$, and $r_{21}=\frac{V_A}{I_{o1}}\dots r_{2M}=\frac{V_A}{I_{oM}}$,
where $V_A$ is the Early voltage and $V_T$ is the thermal voltage. 
Applying Kirchhoff's current law to the small signal circuits in Fig. \ref{fig:small_signal} yields:
\begin{equation}\label{eq:wta1}
\left\{
\begin{aligned}
    &i_{z1} = v_1\frac{I_{z1}}{V_A}+v_c\frac{I_{z1}}{V_T}\\
    &v_j\frac{I_{zj}}{V_A} = -v_c\frac{I_{zj}}{V_T},\quad\forall j\in [2, M]\\
    &\sum_{i=1}^{M}[\frac{I_{oi}}{V_T}(v_i-v_c)+v_c\frac{I_{oi}}{V_A}]
\end{aligned}
\right.
\end{equation}
Given that Early voltage $V_A >> V_T = kT/q$, solving Eq.~\ref{eq:wta1} yields:
\begin{equation}\label{eq:wta2}
\begin{aligned}
    \frac{dV_1}{dI_{z1}} &= \frac{v_1}{i_{z1}} = \frac{1}{I_{z1}}(V_T+V_A(1-\frac{I_{o1}}{I_c}))\\
    \frac{dV_j}{dI_{z1}} &= \frac{v_j}{i_{z1}} = \frac{-1}{I_{z1}}V_A(\frac{I_{o1}}{I_c})
\end{aligned}
\end{equation}
where $I_c=\sum_{j=1}^{M}I_{oj}$.
The $j$th current $I_{oj}$ (see Fig.~\ref{fig:overall}(c)) of the output transistor operating in the subthreshold region can be expressed as:
\begin{equation}\label{eq:wta3}
\begin{aligned}
    I_{oj} = I_oexp((V_j-V_c)/V_T) 
\end{aligned}
\end{equation}
then the term in Eq.~\ref{eq:wta2} 
\begin{equation}\label{eq:wta4}
    \begin{aligned}
        \frac{I_{o1}}{I_c} &= \frac{1}{1+\sum_{j=2}^{M}exp((V_j-V_1)/V_T)}
    \end{aligned}
\end{equation}
Substituting Eq.~\ref{eq:wta4} into Eq.~\ref{eq:wta2} obtains:
\begin{equation}\label{eq:wta5}
    \begin{aligned}
    \frac{dV_1}{dI_{z1}} &= \frac{1}{I_{z1}}(V_T+V_A(1-\frac{1}{1+\sum_{j=2}^{M}exp((V_j-V_1)/V_T)}))\\
    \frac{dV_j}{dI_{z1}} &= -V_A\frac{1}{I_{z1}}(\frac{1}{1+\sum_{j=2}^{M}exp((V_j-V_1)/V_T)}))
    \end{aligned}
\end{equation}
Given that the initial input currents $I_{z1} = I_{z2} = \dots = I_{zM} = I_m$, the initial gate voltage of $T_{21}, \dots,T_{2M}$ should be identical, i.e., $V_m$. 
Eq.~\ref{eq:wta5} can be simplified as below:
\begin{equation}\label{eq:two_region}
    \begin{aligned}
    &\frac{dV_1}{dI_{z1}} = \frac{1}{I_{z1}}(V_T+V_A)\text{ and }
    \frac{dV_j}{dI_{z1}} = 0, \quad\text{ when } V_j-V_1\gg V_T 
    \\
    &\frac{dV_1}{dI_{z1}} = \frac{V_T}{I_{z1}}\text{ and }
    \frac{dV_j}{dI_{z1}} = -\frac{V_A}{I_{z1}}, \quad\text{ when } V_j-V_1\ll V_T
    \end{aligned}
\end{equation}
It can be seen that with $V_j-V_1\gg V_T$ and $V_j-V_1\ll V_T$, the dynamics of the output transistors with the input current are independent of the number of input rails M.
When $V_j\approx V_1 \approx V_m$, Eq.~\ref{eq:wta5} simplifies to:
\begin{equation}
\label{eq:wta7}
    \begin{aligned}
    &\frac{dV1}{dI_{z1}}=\frac{M-1}{M}\frac{V_A}{I_{z1}}\\ &\frac{dV2}{dI_{z1}}=\frac{-1}{M}\frac{V_A}{I_{z1}}
    \end{aligned}
\end{equation}
In 2-rail-input WTA, $V_1$ is a linear function of $I_{z1}$ with a slope of $\frac{V_A}{2I_{z1}}$ \cite{wta_derivation}, and in an M-rail-input WTA, $V_1$ is a linear function of $I_{z1}$ with a slope of $\frac{M-1}{M}\frac{V_A}{I_{z1}}$ as shown in Eq.~\ref{eq:wta7}. 
As the number of input rails scales up, the winner's behavior w.r.t. the dynamics of the output transistor only differs by a constant factor. While for losers $j\in[2, M]$, as the input rail number increases, the behavior w.r.t. the dynamics differ by $\frac{1}{M}$. 
To this end,  the impact of the number of input rails on the winner's output is negligible, which can be seen from Fig.~\ref{fig:class_dimension}(a), where the latency of {\cosi} changes little with increasing number of input rails, i.e., number of class vectors. 
\section{Evaluation}

\label{sec:eval}
In this section, we first evaluate {\cosi} in terms of energy and latency at the array level. We then investigate the scalability and robustness of {\cosi} upon device variations. 
We finally benchmark {\cosi} for binary HDC inference and compare it with a GPU implementation. We have simulated all the {\cosi} components at the circuit-level with Cadence Spectre.
The write voltage for the 1FeFET1R CAM is $\pm 4V$. The 45nm PTM high-performance model is adopted for CMOS transistors~\cite{PTM}, and the Preisach model \cite{ni2018circuit} is used for FeFET. The array wordlength is 1024 bits. The search delay is measured from the beginning of the search operation when the FeFET memory arrays are activated until the WTA circuit generates the output. Besides, the search delay is measured under the worst case, where two non-identical stored vectors are closest to each other, i.e., they only differ by 1 bit at the denominator, and the resulted squared cosine similarities are $cos^2\theta=1/4$ and $1/5$, respectively.

\subsection{Array-Level Evaluation}

\begin{figure}
    \centering
    \includegraphics[width=\linewidth]{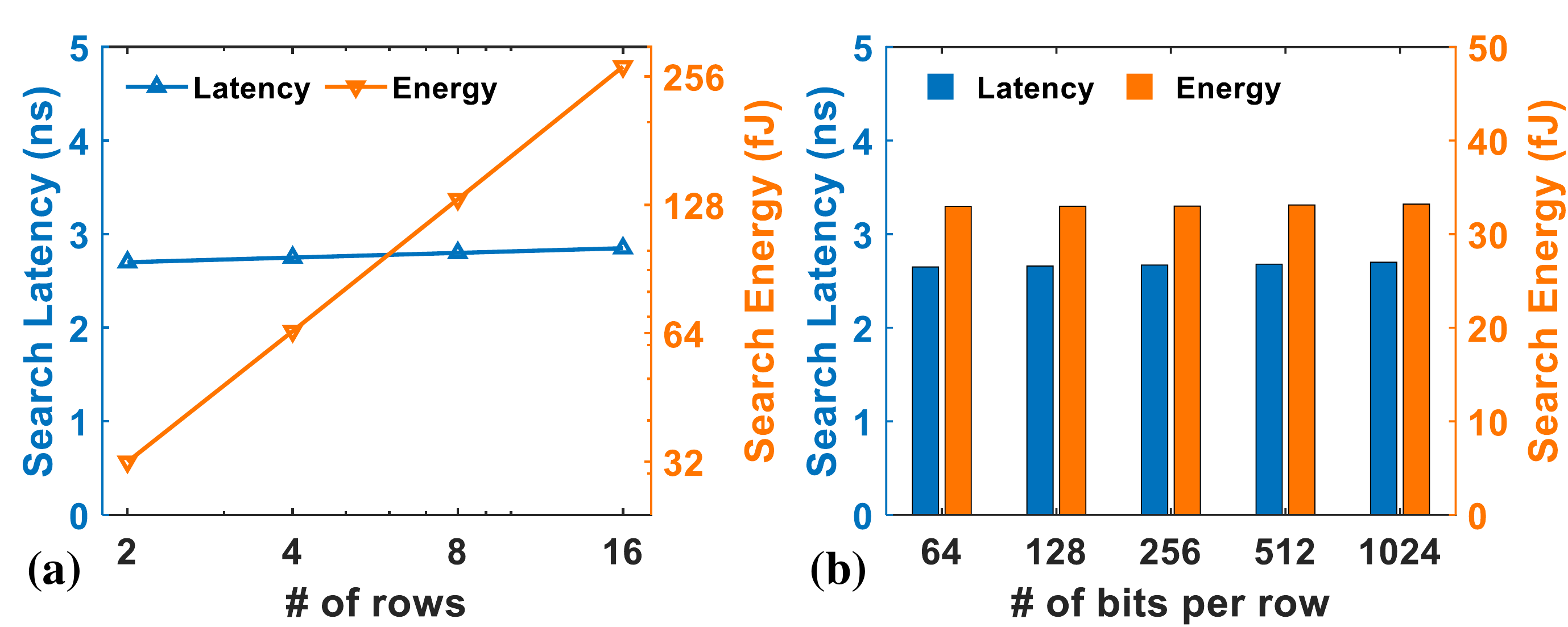}
    \caption{
    Search energy and delay of  {\cosi} with (a) varying number of rows/vectors (1024 bits per row), and (b) varying number of dimensions, respectively.}
    \label{fig:class_dimension}
\end{figure}
\label{sec:array_result}

Fig.~\ref{fig:class_dimension}(a) shows the latency and energy trends of {\cosi} in terms of the number of words in the memory array. 
It can be seen that the increasing class vectors participating the NN search of {\cosi} have negligible impacts on the latency, aligning with the discussion in Sec.~\ref{sec:wta_scale}.
The search energy of {\cosi} mainly consists of two parts: the WTA circuit along with its amplification current mirrors consuming up to $56\%$ of the total energy, and the squared cosine translinear circuits along with their associated current mirrors taking around $43\%$. 
As the number of classes increases, i.e., the number of current paths of WTA circuit increases, the total search energy grows linearly. This is due to the fact that the increasing number of the WTA circuit branches introduces more input and output currents provided by the supply rails.


In addition to varying the number of rows (i.e., classes) within the FeFET memory arrays, we also investigate the scalability of {\cosi} by varying the number of bits in a word (i.e., dimensions) in terms of energy and latency metrics.
As pointed out in Sec. \ref{sec:translinear}, we maintain the correct functionality of translinear circuit and thus {\cosi} by tuning the resistor within the 1FeFET1R structure in the FeFET memory arrays.
Fig. \ref{fig:class_dimension}(b) reports the search energy and latency of {\cosi} with varying number of bits per word in the memory arrays. 
As can be seen, the latency and search energy of {\cosi} has negligible change when increasing wordlength from 64 to 1024, as the total current provided by the supply rails is kept the same based on the tuning method discussed in Sec. \ref{sec:translinear}. 

\begin{figure}
    \includegraphics[width=1\linewidth]{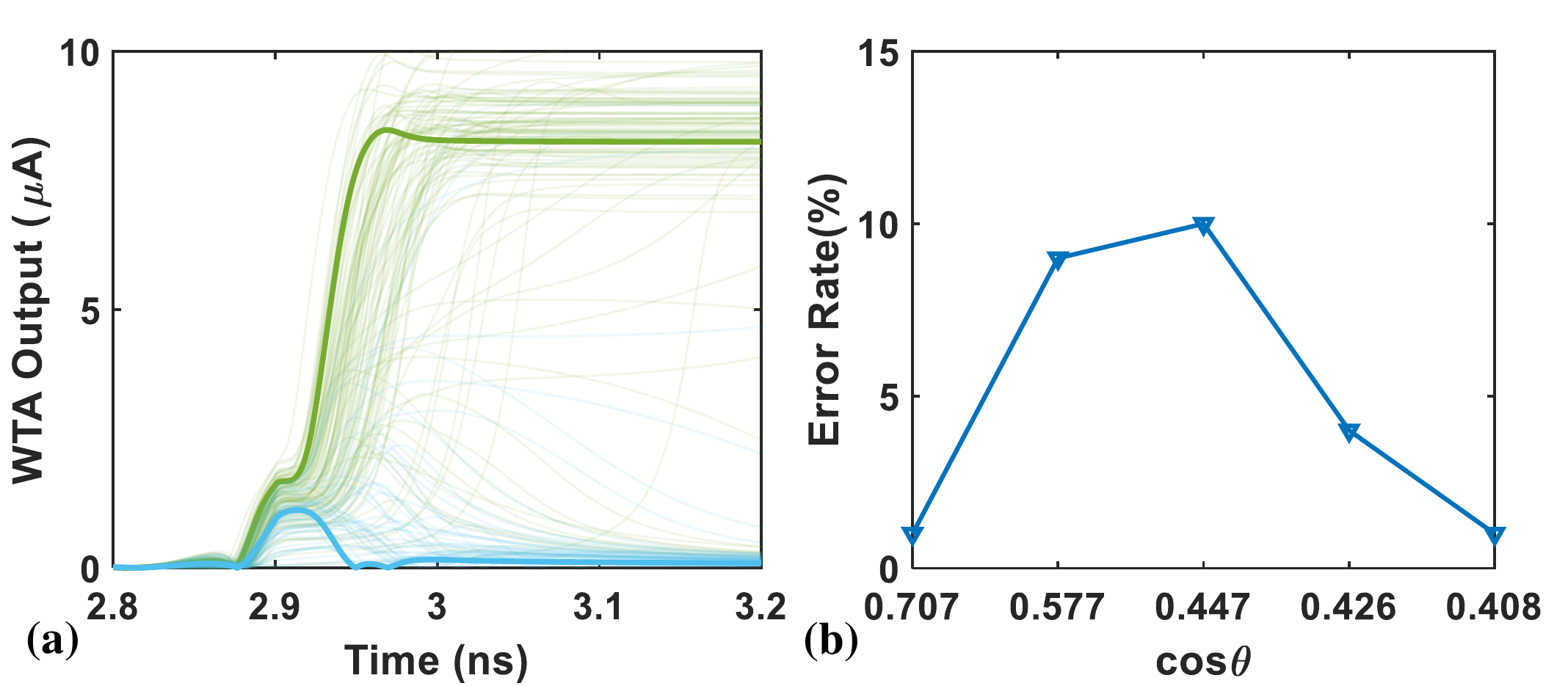}
    \caption{Monte Carlo simulations considering all device-to-device variations: (a) The output waveforms of {\cosi} in the worst-case, achieving 90\% accuracy. (b) The error rates of different cosine similarity outputs with an output $cos\theta=0.5$.}
    \label{fig:MC}
\end{figure}

\begin{figure*}
    \centering
    \includegraphics[width=1\linewidth]{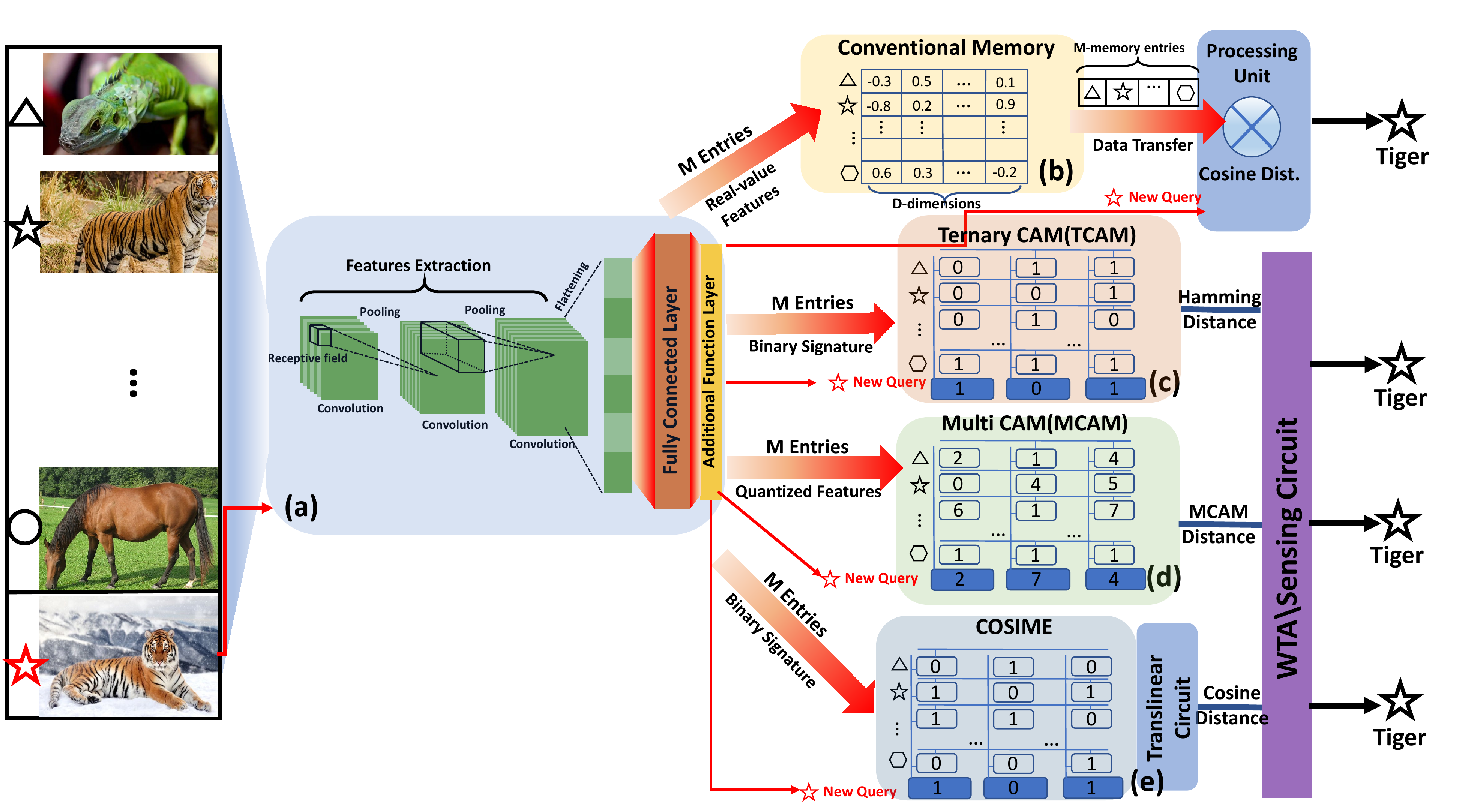}
    \caption{Different AM based implementation along with (a) the feature extractor such as CNN \cite{Ni_2019} and HDC \cite{Ni2022algorithm}. 
    The additional function layer (AFL) following the feature extraction performs local sensitive hashing (LSH) \cite{Ni_2019} or quantization \cite{accuracy}. (b) Conventional memory such as DRAM.   (c) FeFET based AM, e.g. \cite{Ni_2019}, implementing approximate search in Hamming distance. (d) MCAM in \cite{accuracy}, performing NN search in MCAM distance. (e) {\cosi},  performing CSS.}
    \label{fig:high_level}
\end{figure*}

Here we also validate the robustness of  {\cosi} design upon device variability. 
The device-to-device variability of the FeFETs is extracted from \cite{soliman2020ultra}, i.e.,  $\sigma_{LVT}=54mV$ for low-$V_{TH}$ state and
$\sigma_{HVT}=82mV$ for high-$V_{TH}$ state. 
The variability of the resistor in the 1FeFET1R cell is extracted from \cite{1F1R_2021}, i.e., 8\%. The MOSFET device is assumed with 10\% size and 10\% $V_{TH}$ variations, and the supply voltage is assumed with 10\% variation.
Fig.~\ref{fig:MC}(a) shows the output waveforms of {\cosi} for 100 Monte Carlo simulations.
The array-level results indicate a 90\% search accuracy of {\cosi}  with similarity threshold $\cos\theta=0.5$ even in the worst case\footnote{The worst case refers to that the cosine values corresponding to the WTA outputs are $cos\theta = 1/2$ and $1/\sqrt{5}$, respectively. In this case, the two stored vectors differ by 1 bit, which is the harshest situation for WTA circuit to distinguish the corresponding array currents. Such assumption is pessimistic.}. 
Fig.~\ref{fig:MC}(b) shows the array-level search function error rates of {\cosi} generating different cosine similarity values when one entry of the array generates an output corresponding to $cos\theta=0.5$. 
It can be seen from Fig.~\ref{fig:MC}(b) that as the cosine similarities of two stored words with the input query get closer, the error rate of {\cosi} increases, due to the closer current inputs to the WTA circuit.
Yet the maximum error rate is $\approx 10\%$, which would have minimal impact on the application-level accuracy for many machine learning and neuromorphic applications, such as HDC \cite{Imani_2017,Geethan2020,Geethan2021}.


Fig. \ref{fig:high_level} demonstrates and compares different types of AMs implementing various distance metric calculations between the input query and stored vectors for neural network or HDC models. 
Following the feature extractor and additional function layer as shown in Fig. \ref{fig:high_level}(a),
employing conventional memory (Fig. \ref{fig:high_level}(b)) such as DRAM to support NN search incurs significant data movement overhead as all the memory entries need to be sequentially transferred from the memory unit to the processing unit to calculate the cosine similarity between the input query and stored vectors. 
FeFET based AM in \cite{Ni_2019} (Fig. \ref{fig:high_level}(c)) deploys a 2FeFET TCAM array to implement approximate search in terms of Hamming distance between the input query and stored vectors.
Fig. \ref{fig:high_level}(d) from \cite{accuracy} exploits the multi-bit characteristic of FeFET to build an AM implementing NN search in a novel distance metric (i.e., MCAM distance).
Compared with prior AM designs, 
{\cosi} (Fig. \ref{fig:high_level}(e)) achieves superior search energy, performance and area overhead, while still maintaining high accuracy as an AM implementing NN search in cosine similarity distance metric.


TABLE~\ref{tab:comparison} summarizes the distance metric, search energy per bit and latency of different AM designs.
The results show that {\cosi} offers 90.5$\times$ more energy efficiency and 333$\times$ less latency than the approximated CSS design from \cite{Geethan2021}. 
In comparision with the existing AMs with Hamming distance \cite{Imani_2017} or recently proposed Euclidean distance \cite{Kazemi_2021}, {\cosi} is also superior in terms of search energy and latency. 
The significant improvements of {\cosi} over the counterpart approximated CSS design mainly benefit from the following aspects: (1) the advantages of FeFET in read/write energy \cite{Ni_2019}; (2) the 1FeFET1R structure limiting the conducting current  within {\cosi}, which improves the energy efficiency and functionality; and (3) relatively simple analog circuits in {\cosi} compared with the capacitor and analog-to-digital converter (ADC) in \cite{Geethan2021}.
Moreover, TABLE~\ref{tab:comparison} demonstrates the area overheads of different AMs. 
Both the A-HAM in \cite{Imani_2017} and the E$^2$-MCAM in \cite{Kazemi_2021} consume high area overhead since a tree-based loser-take-all (LTA) circuitry and sufficiently large flash cells supporting the 3-bit storage are used, respectively. 
The approximated CSS design in \cite{Geethan2021} consumes $1.31\times$ area overhead than {\cosi} since it adopts ADC for its RRAM readout. 
On the contrary, {\cosi} exhibits ultra-low area overhead since: (1) the analog peripherals of {\cosi} arrays consume much less area than the peripherals of other designs; (2) ultra-compact 1FeFET1R structure has been successfully demonstrated  without consuming extra area overhead \cite{1F1R_2021}.

\begin{table*}[t!]
\centering
\caption{Comparison of Existing AMs with Different Distance Metrics}
\label{tab:comparison}
\resizebox{2\columnwidth}{!}{
\begin{tabular}{c|ccccccc}
\toprule
\textbf{Memory} & \textbf{Technology}& \textbf{Metric} & \textbf{Search Energy per bit$ (fJ)$} & \textbf{Latency}$ (ns)$&\textbf{Area$^*$}$ (mm^2)$ & Process (nm)   \\ \midrule
\textbf{A-HAM}~\cite{Imani_2017}&RRAM&Hamming&$0.20 (\times0.7)$&$8.92 (\times2.9)$&$0.524 (\times 26.5)$&45
\\
\textbf{FeFET TCAM}~\cite{Ni_2019}&FeFET&Hamming&$0.40 (\times 1.4)$&$0.36 (\times 0.12)$&$0.010^\S (\times 0.51)$&45

\\
\textbf{E$^2$-MCAM$^\star$ (1.5 V)}~\cite{Kazemi_2021}&Flash&Euclidean$^2$&$0.56 (\times1.95)$&$5.85 (\times1.95)$&$0.192 (\times 9.7)$&55

\\
\textbf{Approx. Cosine} \cite{Geethan2021}&RRAM&Approx. Cosine&$25.9 (\times90.5)$&$1000 (\times333)$&$0.026^\P (\times 1.31)$&90/65$^\dagger$
\\
\textbf{{\cosi} (this work)}& FeFET&\textbf{Cosine} & $\mathbf{0.286 (\times1)}$ & $\mathbf{3 (\times1)}$&$\mathbf{0.0198} (\times 1)$ &\textbf{45}
\\
\bottomrule

\end{tabular}
}
\flushleft
*: Assuming $256\times256$ array size.
$\S$: Area associated with sensing is not included.
$\P$: Area is estimated via Neurosim3.0 \cite{chen2017neurosim+} and scaled to 45nm technology for fair comparison.
$\star$: E$^2$-MCAM stores 3 bits per cell for search, and the sensing circuitry energy is not included.
$\dagger$: NVM is based on 90nm CMOS while digital peripherals are based on 65m.

\end{table*}

\vspace{-1em}
\subsection{Case Study: Hyperdimensional Classification}
\label{sec:benchmarking}

\begin{figure*}
    \centering
    \includegraphics[width=1\linewidth]{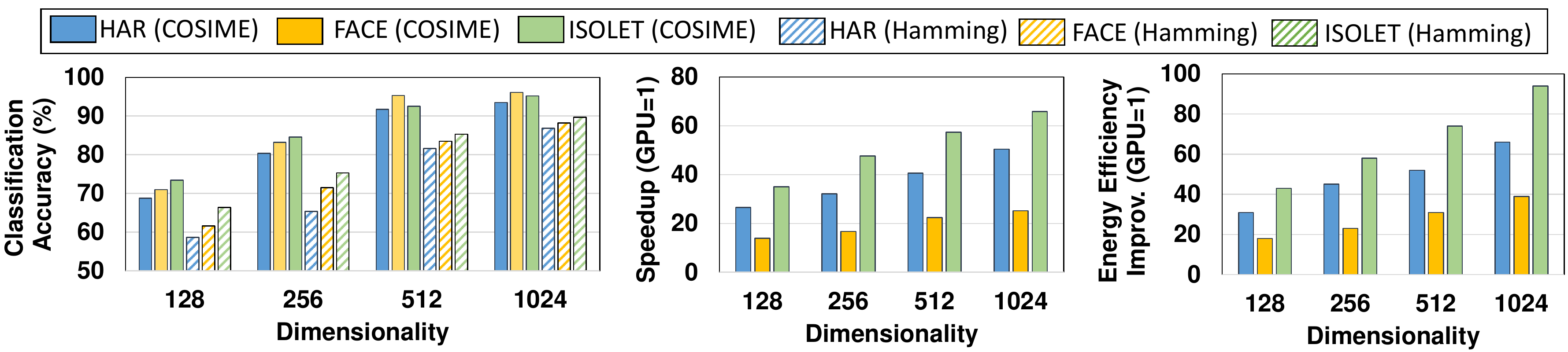}
    \caption{(a) Classification accuracy of HDC using the proposed {\cosi} and Hamming distance as similarity metric. (b) Computation speedup and (c) energy efficiency improvement of {\cosi} compared to GPU.}
    \label{fig:HDC}
\end{figure*}

To validate the effectiveness of {\cosi} as an AM at application level, we benchmark our proposed {\cosi} array in the context of HDC models for classification as a case study.
HDC is based on the understanding that the brain computes with patterns of neural activity that are not readily associated with numbers, and has been proven as effective for many cognitive tasks, such as object tracking~\cite{object_tracking2015}, speech recognition~\cite{ speech2}, image classification~\cite{image1, image2}, etc. 
Due to the size of the brain's circuits, neural patterns can be modeled with hypervectors~\cite{kanerva2009hyperdimensional}. 
HDC builds upon a well-defined set of operations with random hypervectors, is extremely robust upon failures, and offers a computational paradigm that is easily applied to learning problems~\cite{ hernandez2021onlinehd}. 
For HDC classification, the first step is to encode data into high-dimensional space. Then, HDC performs a learning task over encoded data by performing a single-pass training. 
The training generates a hypervector representing each class. 
During the inference phase, as an input query comes which contains the sample data to be classified, it is searched across the stored class hypervectors for the closest one in terms of cosine similarity.
A class with the highest similarity to the query is selected as the inference prediction.  
HDC uses cosine similarity as an ideal distance metric, while prior work approximated with Hamming distance for easier hardware implementation. 
However, this approximation often results in accuracy loss.

Here, we evaluate HDC classification accuracy and efficiency over three large-scale data sets given  in Table~\ref{tab:benchmark}. Fig.~\ref{fig:HDC}(a) shows the HDC classification accuracy when the dimensionality of hypervectors varies from $D=256$ to $D=1k$. 
The results are reported using our proposed {\cosi} (cosine similarity) and Hamming distance~\cite{imani2019bric} as the similarity metrics. 
The results show that HDC achieves maximum accuracy with dimensionality $D=1k$. 
Reducing this dimensionality to 512 and 256 results in 1.7\% and 12.2\% accuracy loss. 
Our evaluation also indicates that by using cosine similarity for distance metric HDC achieves significantly higher accuracy (on average 7\%) as compared to Hamming distance metric. Such observation is consistent with Fig. \ref{fig:accuracy}, and demonstrates the benefits of {\cosi}, which implements CSS for machine learning and HDC tasks. 
Regarding the errors induced by {\cosi} circuits, \cite{Imani_2017,Geethan2020,Geethan2021} have shown that the HDC classification is able to achieve negligible accuracy loss compared with the original accuracy with up to 20\% error rate in the AM. 
Therefore,  {\cosi}, as an AM for HDC, is robust to the device variation even considering the worst case for the HDC classification, as the maximum error rate $\approx10\%$, which is below the HDC error tolerance.

In HDC, the associative search dominates both the training and inference phases (e.g., taking over 90\% of training time~\cite{imani2021revisiting}). Fig.~\ref{fig:HDC}(b), (c) show the energy efficiency improvement and execution time speedup of associative search running on {\cosi} over an NVIDIA 1080 GPU. 
Our results indicate that {\cosi} provides higher speedup and energy efficiency in higher dimensionality. For example, {\cosi} achieves 47.1$\times$ faster and 98.5$\times$ higher energy efficiency on average than GPU with $D=1k$ dimensions. 
{\cosi} provides higher benefits for applications with more classes. 
For example, ISOLET, which has the highest number of classes (see Table~\ref{tab:benchmark}), receives the highest speedup and energy efficiency compared to the GPU implementation. 
This efficiency comes from (1) the capability of  {\cosi} based AM to enable fast and parallel search operations and (2) addressing data movement issues by eliminating data access to off-chip memory.

\begin{table}[t!]

\centering
\caption{Datasets ($n$: feature size, $K$: number of classes)}
\label{tab:benchmark}
\resizebox{1\columnwidth}{!}{
\begin{tabular}{c|cccccc}
\toprule
& $n$ & $K$ & \shortstack{\textbf{Train}\\ \textbf{Size}} & \shortstack{\textbf{Test}\\ \textbf{Size}} & \textbf{Description}             \\ \midrule
\textbf{UCIHAR} & 561               & 12          & 6,213                & 1,554               & Activity Recognition\cite{anguita2012human} \\ 
\textbf{FACE}   & 608               & 2         & 522,441             & 2,494              & Face Recognition\cite{angelova2005pruning}\\ 
\textbf{ISOLET} & 617               & 26         & 6,238               & 1,559              & Voice Recognition~\cite{Isolet}              \\ \bottomrule

\end{tabular}
}
\end{table}


\section{Conclusion}
\label{sec:conclusion}
Hardware acceleration for CSS is important for edge intelligence and AI models.
In this paper, we propose, for the first time, {\cosi}, a FeFET based AM that performs CSS in-memory. 
{\cosi} consists of compact FeFET memory arrays for dot product and squared $L_2$ norm operations, translinear circuits for squaring and division, and the WTA circuit for NN search.
The functionality, scalability and robustness of {\cosi} have been validated. The energy and latency results of {\cosi} at the array level  indicate $90.5\times$ and $333\times$ improvements over the state-of-the-art approximated CSS design, respectively. 
HDC application benchmarking suggests that {\cosi} achieves $47.1\times$ speedup and $98.5\times$ energy efficiency improvement over an GPU implementation. 
Note that the proposed COSIME design is not limited to FeFET technology, but is rather general and can be applied for other NVMs with access transistors. 
This is because the peripheral circuitry of {\cosi} is largely independent of the NVM array as long as the array output currents are within the sensing range.
Therefore, {\cosi} paves a promising way towards efficient CiM designs for CSS in data-intensive applications.

\section*{Acknowledgements}

This work was partially supported by Zhejiang Provincial Key R\&D program (2020c01052, 2022c01232), NSF (LQ21F040006, LD21F040003), NSFC (62104213, 92164203), and Zhejiang Lab (2021MD0AB02). Hu and Niemier were supported in part by ASCENT, one of six centers in JUMP, a Semiconductor Research Corporation (SRC) program sponsored by DARPA.

\bibliographystyle{ieeetr}
\bibliography{bib}
\end{document}